\documentclass[a4paper,twoside]{article}
\usepackage[utf8]{inputenc}
\usepackage[T1]{fontenc}
\usepackage{microtype}
\usepackage{epsfig}
\usepackage{subfigure}
\usepackage{calc}
\usepackage{amssymb}
\usepackage{amstext}
\usepackage{amsmath}
\usepackage{amsthm}
\usepackage{multicol}
\usepackage{pslatex}
\def\tightlist{}
\usepackage[hyphens,spaces,obeyspaces]{url}
\usepackage{hyperref}
\usepackage{siunitx}
\usepackage[british]{babel}
\usepackage[]{marginnote}
\usepackage{csquotes}

\usepackage[%
  style=authoryear,
  backend=biber,
  url=false,
  doi=false,
  isbn=false,
  firstinits=true
]{biblatex}
\DeclareLanguageMapping{british}{british-apa} 
\newcommand{\landau}[1]{\ensuremath{\mathcal{O}\left(#1\right)}}

\usepackage{SCITEPRESS}     

\begin{document}

\title{Distributed Protocols at the Rescue for Trustworthy Online Voting  \subtitle{} }

\author{\authorname{Robert Riemann\sup{1} and Stéphane Grumbach\sup{1}}
\affiliation{\sup{1}Inria Grenoble Rhône-Alpes, Lyon, France}
\email{\{robert.riemann, stephane.grumbach\}@inria.fr}}


\abstract{While online services emerge in all areas of life, the voting procedure
in many democracies remains paper-based as the security of current
online voting technology is highly disputed. We address the issue of
trustworthy online voting protocols and recall therefore their security
concepts with its trust assumptions. Inspired by the Bitcoin protocol,
the prospects of distributed online voting protocols are analysed. No
trusted authority is assumed to ensure ballot secrecy. Further, the
integrity of the voting is enforced by all voters themselves and without
a weakest link, the protocol becomes more robust. We introduce a
taxonomy of notions of distribution in online voting protocols that we
apply on selected online voting protocols. Accordingly, blockchain-based
protocols seem to be promising for online voting due to their similarity
with paper-based protocols.}

\onecolumn \maketitle \normalsize \vfill

\section{VOTING IN THE DIGITAL AGE}\label{voting-in-the-digital-age}

\noindent
Online services have emerged during the last decade in most areas of
every day life. Online banking, online auctions, online shopping, online
learning and online dating, to name a few, have become serious
alternatives to their offline counterparts for the additional
convenience they provide. Most online services are available 24 hours a
day 7 days a week and can be accessed from any place with an internet
connection. In particular, online social networks gained a lot of
momentum. Many people are permanently connected to communicate with
friends and colleagues. The social network Facebook, for instance,
reports on \num{1.71}~billion monthly active users as of June~2016 which
is more than \SI{20}{\percent} of the world population. The adoption of
online services increases further with the global dissemination of
internet-enabled devices.

In these circumstances of pervasive online services, the paper-based
voting procedure used to carry out elections in many democracies appears
like a legacy from the past. This is even more the case for the
generation of Digital Natives that is used to the convenience offered by
online services.

Paper-based voting requires voters to cast their ballot within a given
time frame at a particular location. In contrast, remote electronic
voting, in short online voting, allows voters to cast their ballot from
home or any other place. The time and effort to vote decreases.
Consequently, the overall voter turnout is expected to increase
\autocite{carter2012}. Moreover, the ballot casting is carried out using
a special software which improves the voter convenience as the ballot
presentation considers for instance the voter's spoken language and
possible disabilities.

The voter convenience is especially important if voters are often asked
to vote, e.g.~for elements of direct democracy such as referenda. Not
only doubts in the integrity of the voting outcome, but also a low voter
turnout can negatively impact the legitimacy of the result.

Despite the advantages, only few countries employ online voting for
general elections, e.g.~Estonia \autocite{ulle2006}, Canada
\autocite{goodman2014} and Australia \autocite{brightwell2015}, and for
those, security concerns have been addressed repeatedly by the
scientific community \autocites{ivoting-ccs2014}{halderman2015}. Other
countries have abandoned online voting trials or banned online voting
for their insufficient security, as it happened in Germany. In fact, the
provision of the same set of security properties as known from
paper-based voting for online voting proves to be challenging with no
universal solution, but with different potential concessions.

In this paper, we address the issue of trustworthy online voting
protocols with regard to the potential of distributed protocols.
Starting from paper-based voting with its broadly accepted
trustworthiness (Sec.~\ref{sec:paper-voting}), we give a high-level
review of the principle security concepts of online voting protocols and
deliver an overview of their trust assumptions
(Sec.~\ref{sec:online-voting-concepts}). The capacity of authorities and
of the voters defined by the protocols are hereby of primary concern. We
show that the trustworthiness in those approaches is limited due to the
concentration of power and argue hereinafter that distributed protocols
are promising to increase the trustworthiness due to their similarity
with paper-based protocols. Bitcoin, an electronic currency with no
trusted authority, is introduced in
Sec.~\ref{sec:distributed-protocols}. Moreover, the development towards
online voting is compared with those to Bitcoin and BitTorrent.
Sec.~\ref{sec:voter-empowerment} is dedicated to the prospects of
distributed protocols for trustworthy online voting. Then in
Sec.~\ref{sec:towards-distributed-online-voting}, different notions of
distribution are introduced in order to analyse a selection of existing
online voting protocols for their degree of distribution. Our conclusion
follows in Sec.~\ref{sec:conclusion}.

Our findings motivate our ongoing research on distributed online voting
protocols with a novel distributed protocol currently in preparation for
publication.

\section{PAPER-BASED VOTING}\label{sec:paper-voting}

\noindent
The paper-based voting protocol used today in many democracies is the
result of a development rooted in Ancient Greece where pebbles were
casted in an urn for votings. Since then, many provisions have been
added to ensure the voting outcome reflects indeed the aggregated choice
of all eligible voters. Neglecting local particularities, we give a
sketch of a paper-based voting protocol to recall afterwards how basic
security properties are implemented.

\paragraph{Preparation Phase}\label{preparation-phase}

The voting station is the dedicated location to receive voters on voting
day and is run by volunteering voters that become thus occasionally
voting officers. For large numbers of voters, the voters are partitioned
by locality to a reasonable number of voting districts with each one
voting station. Every station is provided with undistinguishable voting
ballots and a list of eligible voters issued by the central voter
registry.

\paragraph{Casting Phase}\label{casting-phase}

On arrival at the voting station, voters have to present a proof of
identity in order to proceed to the casting. Once the voter is confirmed
to be eligible, a blank ballot is handed out and the voter list is
annotated accordingly to prevent that voters can get more than one
ballot. Under public supervision, the voter enters alone the voting
booth to fill and fold in secret the paper ballot. Again under public
supervision, the filled ballot is thrown into a transparent ballot box.

\paragraph{Aggregation and Evaluation
Phase}\label{aggregation-and-evaluation-phase}

Once, the casting phase is terminated, ballot boxes are opened and
ballots are tabulated to determine the tally that is published
independently along with the derived voting outcome, e.g.~using the
majority rule.

\paragraph{Verification Phase}\label{verification-phase}

Every voter is allowed to attend all phases to supervise the compliance
with the voting protocol.

\paragraph{Conflicting Security
Properties}\label{conflicting-security-properties}

Paper-based voting resolves the dilemma of voting protocols to ensure
\emph{secrecy of the ballot} and \emph{voter eligibility} at the same
time \autocite{lambrinoudakis2003} in a straight-forward way. Voting
officers and voters control that only eligible voters get one single
paper ballot and that every voter puts only one ballot into the ballot
box. Secrecy of the ballot is provided, because every voter fills and
folds its ballot alone and once deposited in the ballot box, all ballots
are indistinguishable and cannot be linked back to the voter. By
eye-sight, ballots can be followed from its distribution, through its
casting into the transparent ballot box until the tallying of all
ballots. \emph{Verifiability} is realised by the observation of the
physical ballot transport, which is called \emph{chain of custody}. This
verification does not require any special knowledge. Hence, neither
trust in the authority carrying out the voting, nor in employed
technology is imposed.

\section{ONLINE VOTING CONCEPTS}\label{sec:online-voting-concepts}

\noindent
The development of online voting protocols, that preserve the security
properties known from paper-based voting, is proven to be difficult.
Electronic ballots can be cloned with no effort and their physical
transport via wire cannot be observed. That is why special cautions must
be taken to prevent the casting of illegitimate ballots, e.g.~of
ineligible voters or voters that seek to cast more than one ballot.
Various concepts have been considered to ensure secrecy, eligibility and
verifiability of online voting. Nonetheless, the overwhelming majority
of current online voting protocols, as we detail hereafter, are either
lacking properties, so that trust in authorities to carry out essential
tasks must be assumed, or use advanced cryptography, which imposes trust
in technology as expert knowledge cannot be implied.

One can distinguish online voting protocols by the following concepts to
ensure both secrecy of the ballot and eligibility
\autocite{lambrinoudakis2003}:

\paragraph{Trusted Authorities}\label{trusted-authorities}

A very basic approach is to assume trust in all or a subset of the
authorities. Voters transmit their ballot to the authorities using an
authenticated channel that allows to verify the eligibility. Authorities
are trusted to keep ballots confidential and to produce the correct
tally.

\paragraph{Anonymous Voting}\label{anonymous-voting}

Using an authenticated channel, voters acquire from one authority an
eligibility token, e.g.~using blind signatures \autocite{chaum1983},
that cannot be linked to the voter's identity, but allows to verify its
eligibility. Then, voters send both ballot and token through an
anonymous channel to the authority. Technological trust is assumed in
the secrecy ensured by the token. Authorities are trusted to produce the
correct tally.

\paragraph{Random Perturbation}\label{random-perturbation}

Voters send encrypted ballots to a group of authorities that shuffle one
after each other the set of all ballots. Shuffling can be realised using
a secure multi-party computation called Mix-Nets \autocite{chaum1981}.
This way, ballots cannot be linked to voters if at least one authority
is honest. Afterwards, ballots are decrypted to compute the tally.
Technological trust and trust in at least one authority is assumed.

\paragraph{Homomorphic Encryption}\label{homomorphic-encryption}

With homomorphic encryption, encrypted ballots are tallied and decrypted
only afterwards \autocite{benaloh1987thesis}. To prevent early
decryption, \((k,n)\)-threshold cryptography \autocite{pederson1991} is
used, which requires the cooperation of \(k\) out of \(n\) authorities.
The use of cryptography implies technological trust and trust in
authorities to the extent that less than \(k\) out of \(n\) authorities
are dishonest.

\paragraph{Balancing Verifiability and Secrecy of the
Ballot}\label{balancing-verifiability-and-secrecy-of-the-ballot}

Online voting protocols offering \emph{end-to-end verifiability}
\autocite{benaloh2014} allow voters to verify the online voting outcome
using cryptographic proofs. However, the very nature of the mentioned
cryptographic protocols prevents to have both universal verifiability of
the voting outcome and unconditional secrecy of the ballot
\autocite{chevallier-mames2010}. Eventually, authorities have to be
trusted to either respect the secrecy of the ballot or deliver the
correct voting outcome. For a trustworthy voting, it is reasonable to
let all voters choose their trusted authorities. Hence, the number of
authorities is in the order of voters, which is infeasible for large
scale elections with protocols based on the presented cryptographic
concepts, because of an amount of required resources polynomial (or
worse) in the number of authorities. As the distribution of powers in
retrospect seems to be problematic, we consider in the next sections
dedicated distributed protocols.

\section{DISTRIBUTED PROTOCOLS}\label{sec:distributed-protocols}

\begin{figure}[b]
\centering
\includegraphics[width=1\columnwidth]{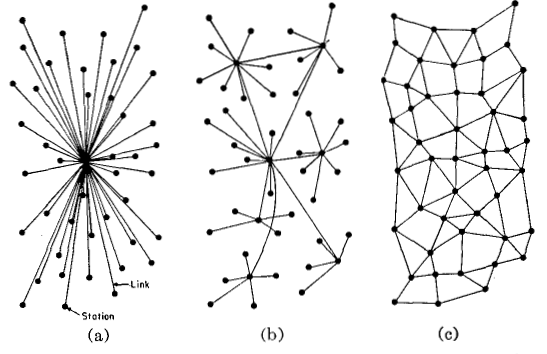}
\caption{(a) centralised, (b) decentralised, (c) distributed systems.
Points represent peers and lines interaction between peers. In (a),
strong specialisation and hierarchy is depicted as one sole peer carries
out a distinct role. The hierarchy becomes more flat in (b) as more
peers serves as intermediaries. In (c), peers are equipotent and
interact with any set of peers, e.g.~within a given
distance.}\label{fig:topology}
\end{figure}

Without consensus on trusted authorities, it is reasonable to omit
authorities altogether if possible and assume instead equipotency of all
voters.

Bitcoin \autocite{bitcoin08} demonstrates the feasibility to find such
protocols without trusted authority. It establishes an electronic
currency without inherent value, prevents its duplication, thus, double
spending, and allows for fast online transactions between peers from
many internet-enabled devices. Therefore, the protocol assumes that the
majority of computing resources is controlled by honest peers.
Technically, Bitcoin can be interpreted as an approach to the situation
in the past, when traders used mostly currencies of inherent value such
as gold, and no trust in online banking systems or authorities like
central banks were assumed \autocite{perezmarco2016}. Hence, after
technological progress allowed a shift from distributed offline to
centralised online systems, Bitcoin presents a distributed online
alternative, c.f.~Fig.~\ref{fig:topology}~(c). A similar analogy is
given with BitTorrent \autocite{cohen2008} that allows its peers to
share their resources to provide information (files) online, when before
central online servers had to be used. The notion of peer
empowerment/democracy in both Bitcoin and BitTorrent offers inspiration
for the development of novel online voting protocols and is further
detailed in the following section.

\section{EMPOWERMENT OF VOTERS}\label{sec:voter-empowerment}

\noindent
In his essay \enquote{Authoritarian and Democratic Technics},
\textcite{mumford1964} explained the arising conflict if technology for
democratic purposes is based on authorities. Indeed, protocols that omit
distinct authorities and instead make provisions for equally privileged,
equipotent voters sharing the same responsibilities, seem to better
reflect the basic democratic concept of equally powerful voters. We
assume further a distributed online voting protocol in which voters are
represented as equipotent peers.

In BitTorrent and Bitcoin, peers cooperate, because they share a common
goal. Peers are providers and consumers at the same time. Applied to
online voting, we find a constellation similar to paper-based voting
where voters serve as voting officers to run the voting station and
carry out the supervision.

Even though not all tasks can be carried out by everyone at the same
time, equipotent peers can replace each other easily. All peers are
responsible to enforce the protocol policy, while byzantine peers with
inconsistent behaviour are often tolerated to a certain degree. These
properties contribute to the robustness of the protocol as there is no
weakest link or bottleneck. Therefore, distributed protocols offer great
potential for resilience against Distributed Denial of Service (DDoS)
attacks.

The peers joining a distributed system, contribute with their own
resources, e.g.~communication capacity, memory and computation power.
The global resources grow and diminish with peers joining or leaving the
system offering potential for great scalability. Further, distributed
protocols allow for ephemeral databases that vanishes entirely when all
peers have left the system, offering new avenues for applications with
sensitive data such as voting. In the case of centralised online voting
protocols, ballots are most often encrypted, but there might be
potential for a malicious decryption at some point in the future, if the
implementation of cryptographic routines turns out to be flawed, secret
keys leak or computing power got sufficiently cheap to brute-force the
encryption.

As every peer has only partial knowledge of the global information, any
unintentional or intentional disclosure is locally bounded. The impact
of an attack, and thus its incentive, is reduced. In contrast, a
disclosure by the authorities of centralised online voting can impact
the secrecy of all casted ballots.

\section{TOWARDS ONLINE VOTING WITHOUT
AUTHORITIES}\label{sec:towards-distributed-online-voting}

\noindent
As presented in Sec.~\ref{sec:online-voting-concepts}, cryptographic
algorithms allow to limit the power of authorities in online voting
protocols. Note, that there is a trade-off. Less trust in authorities is
assumed if authorities are less powerful, but as the protocol complexity
increases, more technological trust is required. Using \emph{Anonymous
Voting} for example, voters do not have to present their ballot in
clear, but have to trust in the properties of the eligibility token.

Further, online voting protocols employing many authorities might assign
authorities either equal roles like in \emph{Random Perturbation} or
\emph{Homomorphic Encryption} where authorities carry out the same
tasks, or different roles. In \emph{Anonymous Voting}, one authority
ensures the eligibility of voters while the other aggregates all ballots
and produces the tally. If all authorities are equipotent and their
number can match the number of voters allowing for an identification
between authorities and voters, the protocol becomes essentially
distributed. In between, there is room for different kinds of partially
distributed protocols that we want to characterise as follows. Note,
that the registration phase shall not be considered in this paper.

\begin{description}
\tightlist
\item[Degree of Specialisation]
ranging from equipotent voters with no specialisation to authorities
with dedicated powers
\item[Topology]
ranging from centralised over decentralised to distributed topologies,
c.f.~Fig.~\ref{fig:topology}
\item[Phase]
Authorities can intervene either not at all, only in few or in all
phases (not distributed).
\end{description}

A protocol shall be called \emph{fully distributed} if the topology is
distributed and voters are equipotent during all phases but the
registration.

Note, that we consider large scale elections for our analysis. First, we
want to qualify paper-based voting as presented in
Sec.~\ref{sec:paper-voting}. There are responsible voting officers, who
can principally exchanged as no special knowledge is required. If we
accept that every voter can become spontaneously voting officer, the
protocol is actually fully distributed during all phases.

\paragraph{Helios}\label{helios}

One of the few published, well-studied online voting protocols with
unconditional end-to-end verifiability is \emph{Helios}
\autocite{adida2008}. While the original version is based on Random
Pertubation (Mix-Nets), the version available online\footnote{\url{https://vote.heliosvoting.org/faq}}
uses Homomorphic Encryption. In both cases trusted authorities are
assumed for threshold decryption requiring communication costs
polynomial in the number of authorities. Hence, the number of
authorities must be small. During the aggregation, ballots are sent to
the authorities for tallying, which corresponds to a centralised
topology. The final step of end-to-end verification is the only
distributed phase, because it can be carried out independently by all
voters once the authorities have published the required material.

While Helios is as such an online voting protocol employing a central
web server to receive ballots and produce the tally, other protocols do
not require such a server as they omit an authority. A selection is
presented hereafter.

\paragraph{Secure Multiparty
Computations}\label{secure-multiparty-computations}

The aim of Secure Multiparty Computations (SMC) is to compute
collectively a joint function over the private inputs of all
participants. The correctness and secrecy properties rely on
cryptography \autocite{yao1982}. It seems to be an appropriate technique
to compute the tally from the individual, secret ballots. However,
several issues render an implementation of a decentralised voting
protocol based on this scheme difficult. As \autocite{gambs2011} points
out, the communication complexity for \(n\) participants is
\(\landau{n^2}\), or in the case of the existence of a trusted party,
\(\landau{n}\). Both render the scheme impractical for large-scale
distributed online voting. Though, SMC is suited for boardroom voting
protocols with only few voters.

Furthermore, the employed cryptography is computationally extensive and
provides often only computational instead of information-theoretic
security.

\begin{table*}[ht]
  \caption{Quality of distribution of selected online voting protocols.}
  \label{tbl:taxonomy}
  \centering
  \begin{tabular}{|l|l|l|l|}
    \hline
    Protocol & Degree of Specialisation & Topology & Distributed Phases \\
    \hline
    Paper-based Voting & none (flexible) & distributed & all \\
    \hline
    Helios, \cite{adida2008} & selected authorities & centralised & verification \\
    \hline
    SPP, \cite{gambs2011} & random authorities & structured, tree & aggregation \\
    \hline
    DPol, \cite{guerraoui2012} & none & structured, ring & all \\
    \hline
    Blockchain-based Voting & none (flexible) & distributed & all \\
    \hline
  \end{tabular}
\end{table*}

\paragraph{Scalable and Secure Aggregation
Protocol}\label{scalable-and-secure-aggregation-protocol}

\begin{figure}[tb]
\centering
\includegraphics[width=1\columnwidth]{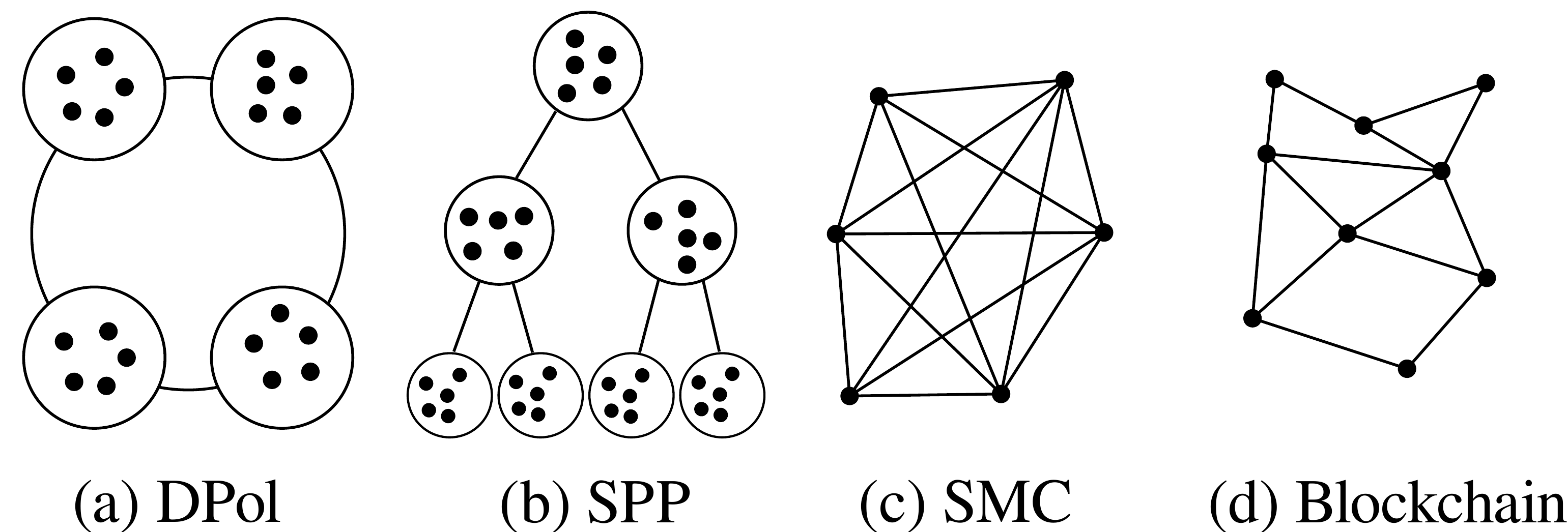}
\caption{Topology of distributed voting protocols. (a) DPol and (b) SPP
have a higher degree of hierarchy than (c) SMC and (d) blockchain-based
protocols due to partition of voters in clusters. This way, the number
of exchanged messages can be reduced at the cost of less equipotent
voters with less flexible roles. Unstructured SMC- and blockchain-based
protocols use a distributed topology allowing for equipotent
voters.}\label{fig:topology-protocols}
\end{figure}

An intermediate solution between a centralised online voting protocol
based on cryptography and an aggregation by the voters is given with
Scalable and Secure Aggregation (SPP) \autocite{gambs2011}. Voters are
grouped using a chord overlay into clusters of equal size in order to
partition potential dishonest voters. The clusters are then arranged in
a binary tree structure (Fig.~\ref{fig:topology-protocols}~(b)). Voters
assigned to the root cluster have to create a key pair for homomorphic
public-key cryptography. Thus, a \((k,n)\)-threshold decryption key pair
is jointly created using a distributed key generation protocol. While
the private key parts are distributed among the members of the root
cluster, the public key is communicated to all child clusters in the
tree.

Every voter encrypts its ballot and adds a non-interactive
zero-knowledge-proof to back the validity of its vote. Encrypted ballots
are gathered from voters in the same cluster to compute an intermediate
aggregate for the cluster. From the tree leaves, aggregates are passed
to all members of the parent cluster, who compute first the sub-tree
aggregate taking into account the own aggregate and those of its child,
and pass the result to all members of the parent cluster. By majority
rule, voters decide in case of diverging results which aggregate has to
be used for the computation.

At some point, the root cluster will possess the encrypted final tally
of all ballots. A union of \(k\) voters will carry out the joint
decryption algorithm to compute the final tally that is subsequently
propagated down the tree to all voters.

Even though all voters are involved in the aggregation procedure, the
parity of knowledge is not given. The voters in the root cluster
correspond effectively to randomly determined, trusted voting officers
in a centralised protocol, that cannot be flexibly exchanged after the
key pair generation.

\paragraph{Decentralised Polling}\label{decentralised-polling}

Another technique to realise distributed online voting is based on
secret sharing schemes. A corresponding protocol known as Decentralised
Polling (DPol) is described in \autocite{guerraoui2012}. \(n\) voters
are grouped to clusters of size \(\sqrt{n}\). The clusters form a ring,
so that each cluster has one preceding and one succeeding cluster
(Fig.~\ref{fig:topology-protocols}~(a)). Every voter gets accorded
\(2k+1\) recipient voters from the succeeding cluster and is itself
recipient to \(2k+1\) voters from the preceding cluster. \(k\) is a
privacy parameter. A ballot for a particular option from a set of size
\(d\) is represented by a vector in \(\{0,1\}^d\). The voter issues then
\(2k+1\) vectors, \(k+1\) towards its choice and \(k\) for others. The
vectors are distributed among the receivers. In the next step, received
vectors are summed up and the result is exchanged with other voters of
the same cluster in order to compute the tally for the preceding
cluster. The tallies are then communicated to the succeeding clusters in
the ring until every cluster is in possession of all the cluster tallies
such that the final tally can be computed by every voter.

It is remarkable, that this scheme provides secrecy of the ballot
without the employment of cryptography. The protocol authors describe
means to determine dishonest voters with high probability without
false-positives and tag them accordingly, e.g.~in a public social
network profile.

Few extensions have been proposed, among them EPol \autocite{hoang2014},
that generalises the DPol voting system in such a manner, that the
actual social graph structure (with some assumptions) can be used and a
ring social net overlay and a perfect square voter number is not a
prerequisite anymore.

Both DPol and EPol provide for equipotent voters. The aggregation is a
joint effort in which all voters are involved. Every voter computes the
final tally and then the voting result. Similar to the SPP protocol,
intermediate aggregates are employed to keep the information on
individual ballots local in the overlay.

\paragraph{Blockchain-based Voting
Systems}\label{blockchain-based-voting-systems}

Online voting based on Bitcoin technology, namely the global consensus
algorithm given by its blockchain, is eagerly anticipated by some
groups. Various prototypes and commercial solutions are or have been
developed\footnote{\raggedright \url{https://github.com/domschiener/publicvotes},
  \url{http://votem.com}, \url{http://www.bitcongress.org},
  \url{http://followmyvote.com}, \url{http://cryptovoter.com},
  \url{http://votosocial.github.io}}. However, the actual protocols are
either lacking essential properties or remain obscure. Scientific
results are very sparse. \textcite{zhao2015} present a protocol for a
binary vote that requires a monetary deposit and a funding of the winner
by all voters. Transactions to the winner and voter payback is
controlled due to contracts enforced by the distributed network which
limits the flexibility of the ballot.

To the author's knowledge, most other approaches to online voting using
Bitcoin are based on so-called \emph{coloured coins} that allow to
associate digital assets to Bitcoin addresses. Consequently, asset
ownership can be traded like the Bitcoin currency and (pseudonymous)
ownership is publicly verifiable by following the previous asset
transactions.

To construct an online voting protocol, every voter must initially own a
coloured coin representing its eligibility. Those coins are then
transferred to a destination that corresponds to e.g.~a candidate.
Transactions are publicly verifiable, so that aggregation and evaluation
can be carried out by every voter. Though, the publicity of all
transactions endangers the secrecy of the ballot, because coins can be
linked back to the voter that has been identified during the
registration phase to receive its coin.

Different protocols have been proposed to provide anonymous Bitcoin
transactions \autocites{miers2013}{ibrahim2016} that are considered to
ensure secrecy of the ballot in blockchain-based online voting. Also
blind signatures may permit to distribute coins anonymously to voters.
In that case, the protocol approaches the Anonymous Voting concept
(c.f.~Sec.~\ref{sec:online-voting-concepts}). The authority carrying out
the aggregation is hereby replaced by the distributed blockchain that
serves as a public add-only bulletin board to log casted ballots.
Consequently, tallying and verification can be carried out by every
voter. The distributed topology is inherited from Bitcoin which uses
gossiping to spread transactions
(Fig.~\ref{fig:topology-protocols}~(d)). Transactions are then gathered
into blocks by voters or third parties that seek to confirm
transactions. Every voter can as its own discretion engage in the
confirmation procedure. There is no specialisation of voters.

It turns out that blockchain-based and paper-based voting are on par
with respect to the notions of distribution and qualify both as fully
distributed, c.f.~Tab.~\ref{tbl:taxonomy}.

\section{CONCLUSION}\label{sec:conclusion}

\noindent
Major votings in 2016, e.g.~the Brexit referendum or the US presidential
elections, demonstrated the importance of a high voter turnout for the
legitimacy of the outcome, especially in the case of tight outcomes.

While online voting is generally considered with much hesitation,
advances in technology are eroding the security of paper-based voting.
Coercion-freeness, in the past a major argument for on-site ballot
casting, is at stake due to omnipresent smart phone cameras. Exit polls
on social media allow voters casting their ballot very late a more
informed choice, which harms the fairness. More and more voters use
early postal voting sacrificing thus their means of verification and the
potential to change their vote last minute. Meanwhile, many online
voting protocols seek to achieve those security properties that
paper-based voting with optional postal voting can ensure less and less.

In this situation, distributed online voting offers a promising
perspective. It does not assume a trusted authority and the integrity of
the voting is enforced by the voters themselves. Without a weakest link,
votings are difficult to interrupt. The damage in case of security
breaches is locally bounded thanks to the distribution of data. Like in
all online voting protocols, voting becomes more convenient, especially
as ballots can be casted from remote. We acknowledge that still trust in
technology or expert knowledge is assumed and leave it as an open issue.

So far, only few distributed online voting protocols have been proposed
and even less are fully distributed. We hope to see more proposals in
the future and plan to contribute to this subject with a novel fully
distributed protocol. This work in progress follows the Anonymous Voting
concept and uses techniques from BitTorrent to create a voter overlay
network and from Bitcoin to ensure verifiability. Similar to DPol,
secrecy is provided by a particular sharing scheme instead of
cryptography. To achieve logarithmique complexity, it is based like SPP
on a tree overlay network. Unlike in SPP, no decryption step is needed.
Hence, all voters are equipotent.

\section*{\uppercase{Acknowledgements}}

\noindent
The authors wish to thank Stéphane Frénot, Damien Reimert and Aurélien
Faravelon for fruitful discussions on distributed voting protocols.

\section*{\uppercase{References}}

\renewcommand*{\bibfont}{\small}
\printbibliography[heading=none]

\vfill
\end{document}